\documentclass{ws-mpla}

\newbox\grsign      \setbox\grsign=\hbox{$>$}
\newdimen\grdimen   \grdimen=\ht\grsign
\newbox\simgreatbox \setbox\simgreatbox=\hbox{\raise.5ex\hbox{$>$}\llap
                        {\lower.5ex\hbox{$\sim$}}}\ht1=\grdimen\dp1=0pt
\newbox\simlessbox  \setbox\simlessbox =\hbox{\raise.5ex\hbox{$<$}\llap
                        {\lower.5ex\hbox{$\sim$}}}\ht2=\grdimen\dp2=0pt
\def\simgreat{\mathrel{\copy\simgreatbox}}
\def\simless {\mathrel{\copy\simlessbox }}

\begin{document}

\markboth{R. Demarco, P. Rosati \& H. C. Ford}
{Panchromatic Studies of Distant Clusters of Galaxies}

%
\catchline{}{}{}{}{}
%

\title{PANCHROMATIC STUDIES OF DISTANT CLUSTERS OF GALAXIES}

\author{\footnotesize Ricardo Demarco and Holland C. Ford}

\address{Department of Physics and Astronomy, Johns Hopkins University, 3400 N. Charles Str.,\\
Baltimore, MD 21218, USA\\
demarco@pha.jhu.edu}

\author{Piero Rosati}

\address{ESO-European Southern Observatory, Karl-Schwarzschild Str. 2,\\
Garching b. M\"unchen, D-85748, Germany
}



\maketitle

\pub{Received 5 May 2005}{}

\begin{abstract}

High redshift ($z \simgreat 1$) clusters are ideal probes to study the
formation and evolution of large scale structures and galaxies in the
universe. A 10-m class ground based telescope, X-ray observatories
(Chandra, XMM-Newton) and HST/ACS are allowing us to perform an
unprecedented study of distant massive clusters of galaxies in the
redshift range $0.84<z<1.3$, selected from X-rays surveys. In this
paper we summarize our results on the structure and dynamics of two of
these clusters derived from imaging and spectroscopic data as well as
our results on the evolution of early-type galaxies.

\end{abstract}

\ccode{PACS Nos.: 98.65.Cw, 98.65.Hb, 98.62.Ai, 98.62.Py, 98.62.Lv, 98.80.Es}

\section{Introduction}\label{intro}

Clusters of galaxies are the largest gravitationally bound structures
known in the universe, constituting a collection of galaxies embedded
in a hot (T $\sim 10^{7}$-$10^{8}$ K) and thin ($n_e \sim 10^{-3}$
cm$^{-3}$) gas in the gravitational potential created by the cluster
dark matter (DM) halo. The velocity dispersion of galaxies in clusters
is typically $\sigma_v \sim$ 800 - 1000 km s$^{-1}$. The scale of a
cluster of galaxies is of the order of a few Mpc, and its mass can 
vary between $10^{14}$ and $10^{15} M_\odot$.

Clusters of galaxies are fundamental objects in modern cosmology.
They represent regions where mass overdensities have reached maximum
amplitudes; consequently, the study of their correlation in space can
be used to estimate the power spectrum of density fluctuations in the
universe. The mass function of distant galaxy clusters is sensitive to
the cosmological parameters defining the geometry and dynamics of the
universe\cite{BC93,BFC97,FBC97}, therefore galaxy clusters can be used
to probe the cosmological model. Moreover, galaxy clusters are ideal
laboratories to study the effects of environment on the formation and
evolution of galaxies, in particular of early-type galaxies, the most
massive galaxies known. Predictions of the spectro-photometric
properties of early-type galaxies in clusters based on hierarchical
models of galaxy formation\cite{K96,KC98a,KC98b} differ significantly
at redshifts $z \simgreat 1$ from models in which ellipticals formed
at high redshift in a monolithic collapse.\cite{ELS62} A detailed
analysis of the internal structure and dynamics of galaxy clusters is
thus essential to understand their formation and evolution as well as
the formation and evolution of their baryon content (galaxies and
gas).

In this context, we have undertaken an ambitious observational program
to obtain high angular resolution imaging with the Advanced Camera for
Surveys (ACS)\cite{F02} on the HST of 7 clusters of galaxies\cite{F04}
with redshift in the range $0.8 < z < 1.3$. The aim of this program is
to study the formation and evolution of clusters and their galaxy
populations. These data are supported by other space and ground based
observations covering a wide range in wavelength, from X-rays to
near-IR. As an example of our program, in this paper we review our
panchromatic study of two of these clusters, which are among the most
distant ($z \simgreat 1$) massive clusters of galaxies known today: RX
J0152.7-1357 ($z=0.84$; hereafter RX J0152 for brevity) and
RDCS1252.9-2927 ($z=1.24$; hereafter RDCS J1252 for
brevity). Multi-wavelength imaging data and optical spectroscopy are
combined to provide detailed information about the structure of these
clusters and the spectrophotometrical properties of their galaxy
populations. Throughout this paper, unless otherwise indicated, we
assume a $\Lambda$CDM cosmology with $H_0$ = 70 $km s^{-1} Mpc^{-1}$,
$\Omega_m$=0.3 and $\Omega_{\Lambda}$ = 0.7.

\section{Physics of galaxy clusters}\label{theory}

Multi-wavelength observations of galaxy clusters are fundamental tools
to examine the physical properties of their components: DM and Baryons
(gas and galaxies). DM contributes $\sim$ 85\% of the cluster mass
while the remaining mass comes mostly from the gas. Galaxies represent
a tiny fraction ($\sim$1\%) of the cluster mass. Optical and near-IR
observations allow us to study the physics of cluster galaxies and
X-ray observations show us the physical properties of the intracluster
medium (ICM) gas.

Clusters of galaxies are among the strongest sources of X-rays with
typical X-ray luminosities of $L_X \sim 10^{43}-10^{45}$ ergs
s$^{-1}$, which makes them detectable at redshifts beyond unity. This
emission is due to thermal Bremsstrahlung from the interaction between
the electrons and ions of the hot ICM. In a local thermodynamical
equilibrium state at temperature $T$, the X-ray emissivity of the ICM,
in units of $W \ m^{-3} \ {\mathrm Hz}^{-1}$, is \cite{BM98}:

\vspace{0.2cm}
\begin{equation}
\label{brems}
\epsilon_\nu(\textbf{r})= 5.44\times 10^{-52} \overline{Z^2} n_e(\textbf{r})n_i(\textbf{r}) T^{-1/2}(\textbf{r}) g^{ff}(Z,T(\textbf{r}),\nu) e^{-h \nu / (k T(\textbf{r}))} \ ,
\end{equation}

\vspace{0.2cm}
\noindent
where $n_e(\textbf{r})$ is the electron number density,
$n_i(\textbf{r})$ is the ion number density and $\overline{Z^2}$ is
the mean-square atomic charge on the ions. The coefficient
$g^{ff}(Z,T(\textbf{r}),\nu)$ is a correcting factor, called the Gaunt
factor.\cite{KL61,KBK75} This equation can be fit to the observed
X-ray spectrum to estimate the ICM temperature and metallicity and,
since Eq. (\ref{brems}) depends on the square of the gas density
($\rho^2_{gas} \propto n^2_e \propto n_e n_i$), a model for
$\rho_{gas}$ can be used to fit the cluster X-ray surface brightness
and obtain the gas distribution. A popular description for the density
profile of an iso-thermal gas in hydrostatic equilibrium is the
$\beta$-model\cite{CF76}:

\vspace{0.2cm}
\begin{equation}
\label{beta-model}
\rho_{gas}(r)=\rho_0 \left\{ 1 + \left(\frac{r}{r_c} \right)^2 \right\}^{-3 \beta / 2} \ .
\end{equation}

\vspace{0.2cm}
\noindent
The parameter $\beta$ is the ratio between kinetic galaxy energy and
thermal gas energy: $\beta = (\mu m_p \sigma^2_v)/(k_B T)$, where
$\sigma_v$ is the galaxy velocity dispersion along the line of
sight. Once the temperature is known and assuming that the cluster gas
is in hydrostatic equilibrium, then the cluster mass distribution is
given by:

\vspace{0.2cm}
\begin{equation}
\label{xmass}
M( < r)=-\frac{k_B r}{\mu m_p G} T \left\{ \frac{d\ log\ \rho_{gas}(r)}{d\ log\ r}+\frac{d\ log\ T(r)}{d\ log\ r} \right\} \ .
\end{equation}
\vspace{0.2cm}

This mass, obtained from X-ray observations, can be compared with other
mass estimates, such as the virial mass and the gravitational lensing
mass. Spectroscopic observations in the optical of cluster galaxies
allow us to measure the galaxy velocity dispersion along the line of
sight, $\sigma_v$. Once the projected cluster galaxy positions are
obtained from the imaging data, the virial mass of the cluster can be
computed as\cite{GGM98}:

\vspace{0.2cm}
\begin{equation}
\label{virmass}
M_V = \frac{3 \pi}{2} \frac{\sigma^2_v}{G} \left( \frac{N(N-1)}{\sum_{i>j} R^{-1}_{ij}} \right) \ ,
\end{equation}

\vspace{0.2cm}
\noindent
where $N$ is the number of galaxies and $R_{ij}$ is the distance
between any pair of galaxies. Since the velocity dispersion $\sigma_v$
is dependent on the dynamical state of the cluster,
Eq. (\ref{virmass}) provides a true estimate of the cluster mass only
if the structure is relaxed. In cases where the cluster structure is
not in virial equilibrium, Eq. (\ref{virmass}) represents an upper
limit to the cluster mass. In contrast, the weak lensing
analysis\cite{M99} offers a way of estimating masses which is
independent of the cluster dynamical state. However, weak lensing
turns out to be sensitive to the distribution of matter along the line
of sight as well as to the distribution of background
sources. Therefore, a good knowledge of these distributions is
required to obtain reliable mass estimates. The strong lensing
technique\cite{M99} is a more robust mass tracer; however, it requires
the existence of gravitational arcs in the cluster field, which are
not always present or detectable, and a knowledge of their redshifts.

Assuming that clusters of galaxies are in a relaxed state and that
gravity is the only dominant driver of evolution, theory shows that
these objects are self-similar structures described by scaling
relations relating mass, temperature and luminosity. In particular, if
the gas fraction in clusters is independent of temperature, then a
relation $L_x \propto T^2$ is expected. Observational evidence shows,
however, that local clusters follow a relation $L_x \propto
T^{\alpha}$, with $\alpha > 2$. Although some analyses indicate a
value of $\alpha$ independent of temperature\cite{MZ98,OP04}, there is
observational evidence that points toward larger values of $\alpha$ on
group scales making the $L_x - T$ relation even steeper at lower
temperatures ($T \simless 1 \ \mathrm{keV}$).\cite{HP00} A departure
from theoretical predictions is also shown by the observed $L_x - M$
and $T - M$ relations.\cite{SPF03,FRB01} The departure from
self-similarity of the $L_x-T$ relation suggests that energy injection
from non-gravitational processes\cite{LPC00,BBP99,TN01,BGW01} or a
temperature dependent gas fraction may be the cause of the departure
from theory, although the latter is likely not the case.\cite{ETB04}
When observing clusters or groups of galaxies at higher redshifts (up
to $z \sim 1$), the behavior of the $L_x - T$ relation shows no major
change.\cite{R04b} This together with a non evolving ICM metallicity
of about 0.3 $Z_{\odot}$\cite{R97,TRE03}, since $z \sim 1$ down to the
local universe, suggest an early formation epoch for the ICM with the
major episode of metal enrichment and gas preheating occurring at $z >
1$. With the development of more advanced techniques, detailed
hydrodynamical simulations\cite{B04} of cluster formation including
galaxy feedback are being performed in order to reproduce the
observations, casting more light on the above issues and deepening our
understanding of cluster formation.

In the optical range, spectroscopic and imaging observations of
cluster galaxies provide us with valuable information about the
dynamics of the cluster itself and the properties of the member
galaxies. Spectrophotometric surveys are essential in determining the
overall galaxy Luminosity Function (LF) of clusters, a powerful tool
for constraining models of galaxy formation and evolution. The LF is
commonly modeled with a Schechter function\cite{S76}:

\begin{equation}
\phi (L) = \frac{\phi^*}{L^*} \left( \frac{L}{L^*} \right)^{- \alpha} e^{- (L/L^*)} \ ,
\label{lf}
\end{equation} 

\noindent
where $L^*$ is the characteristic luminosity of the population of
galaxies and corresponds to the luminosity over which the exponential
term dominates, i.e., the bright end of the LF. The index $\alpha$
gives the slope of the power law behavior of the LF at low
luminosities, and the normalization $\phi^*$ determines the volume
density of sources $n_0$ as $n_0=\int^{\infty}_0 \phi(L) dL=\phi^*
\Gamma(1 - \alpha)$. Observations of the galaxy LF in clusters over a
wide range in redshift ($0 < z \simless 1$)\cite{dPSE99,EJ04} show
compelling evidence for the evolution of bright (massive) galaxies in
clusters that is consistent with pure passive evolution models with
$z_f \simgreat 2$. If merging plays a role in the formation of these
galaxies, these results suggest that any merging activity should take
place at higher redshift. The same conclusion is obtained from the LF
of field galaxies\cite{P03}, showing that massive elliptical galaxies
were already in place at $z \simeq 1$, forming their stars and
assembling their mass at higher redshift. Passive evolution models of
early-type galaxies in clusters also show a good agreement with
photometric observations of cluster galaxies. The color-magnitude (CM)
diagram of clusters shows a well defined red sequence, the locus
formed by the majority of red early-type cluster galaxies. The tight
CM relation observed in local\cite{BLE92,TCB01} and
distant\cite{SED98,R99,vDSH01,SHR02,B03} cluster early-type galaxies
indicates that in clusters these galaxies are formed at redshifts $z
\simgreat 2$, after an initial burst of star formation followed by
pure luminosity evolution, posing difficulties for hierarchical models
that have ``late'' ($z < 2$) formation of the brightest galaxies. In
addition, spectroscopic observations of cluster galaxies allow us to
explore their stellar content, thus providing essential information to
better constrain galaxy evolution models of galaxies and understand
the galaxy-ICM connection during the cluster evolution.

\section{The observational data}

As an example of our cluster program, in this paper we present a
summary of results obtained so far from our multi-wavelength study of
two distant clusters RX J0152 at $z=0.837$\cite{DCSG00} and RDCS J1252
at $z=1.237$\cite{R04b}, both selected from the Rosat Deep Cluster
Survey (RDCS)\cite{RDCN98}. In addition to the ACS data, they have
also been targeted for X-ray (Chandra and XMM-Newton satellites) and
optical and near-IR (ESO NTT and ESO VLT) observations as well as for
optical spectroscopy (ESO VLT). The ground based optical and near-IR
imaging as well as the VLT spectroscopic survey of RX J0152 are
presented in Demarco et al. (2005).\cite{DRL05} ACS observations of
the same cluster are discussed and analysed in Blakeslee et al. (in
preparation), Homeier et al. (2005)\cite{HDR05}, and Jee et
al. (2005)\cite{JWB05} and the X-ray observations are presented in
Maughan et al. (2003).\cite{MJE03} In the case of RDCS J1252, very
deep near-IR imaging with VLT/ISAAC\cite{LRD04} was obtained in very
good seeing ($<$0.5) conditions allowing us to reach a limiting
magnitude (5$\sigma$ detection threshold over a $0''.9$ diameter
aperture) in the Vega system of 25.6 and 24.1 in the J- and K$_s$
-bands respectively. X-ray observations of RDCS J1252 are described in
Rosati et al. (2004)\cite{R04b} and the ACS observations of this
cluster are presented in Blakeslee et al. (2003)\cite{B03} and Holden
et al. (2005).\cite{HBP05}

\begin{table}[h]
\tbl{Physical properties of RX J0152.7-1357 and RDCS
J1252.9-2927. Redshift and velocity dispersion values are from Demarco et al. 2004, 2005 and Demarco et al. 2005, in preparation. Temperatures, X-ray luminosities and metallicities are from Ettori et al. 2004.}
{\begin{tabular}{@{}lccccc@{}} \toprule
Cluster & z & Temperature & X-ray $L_{bol}$ & Rest Frame Vel. Disp. & Metallicity \\
 & & (keV) & $\times 10^{44}$ ({\textrm ergs} {\textrm s}$^{-1}$) & (km s$^{-1}$) & ($Z_{\odot}$) \\
\colrule
RX J0152.7-1357 N & 0.839 & 6.0$^{+1.1}_{-0.7}$ & 10.67$\pm$0.67 & 919$\pm$168 &0.17$^{+0.19}_{-0.16}$ \\
RX J0152.7-1357 S & 0.830 & 6.9$^{+2.9}_{-0.8}$ & 7.73$\pm$0.40  & 737$\pm$126 &$<$0.22 \\
\colrule
RDCS J1252.9-2927 & 1.237 & 6.0$^{+0.7}_{-0.5}$ & 6.6$\pm$1.1 & 766$\pm$89 & 0.36$^{+0.12}_{-0.10}$\\
\botrule
\end{tabular}}
\label{properties}
\end{table}

\section{The galaxy cluster RX J0152 (z=0.837)}

RX J0152 is a massive dynamically young system at $z=0.837$ where two
clusters of galaxies are likely merging. Since its discovery with
ROSAT\cite{DCSG00}, a number of observations in the X-ray domain as
well in the optical have been carried out, confirming its complex
structure and unvirialized state.\cite{MJE03,JWB05,DRL05,HDR05}

\begin{figure}[th]
\centerline{\psfig{file=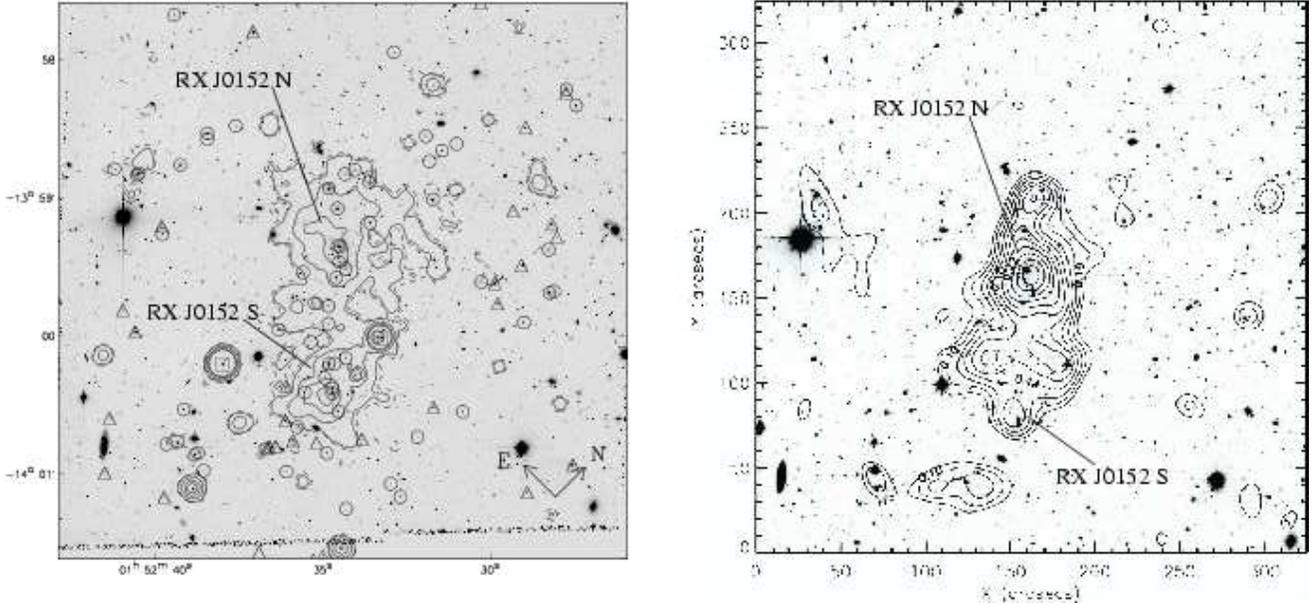,width=8cm,angle=-90}}
\vspace*{8pt}
\caption{RX J0152. {\bf Left:} Chandra X-ray iso-contours overlaid on
the ACS optical image. The circles are spectroscopically confirmed
``passive'' (no emission lines) cluster members, the triangles are
confirmed star-forming galaxies, and the two squares at the centers of
two X-ray point sources are confirmed cluster Seyferts (Demarco et
al. 2005). {\bf Right:} Mass iso-contours from the weak lensing
analysis (Jee et al. 2005) overlaid on the ACS optical image (Figure
taken from Jee et al. [2005]).}
\label{c0152}
\end{figure}

\subsection{RX J0152: Cluster structure and dynamics}

Fig. \ref{c0152} shows the distribution of Baryons and DM in RX
J0152. The X-ray data from Chandra (Fig. \ref{c0152}-Left) clearly
shows that RX J0152 is composed of two extended central substructures;
the centroids of these two substructures (subclusters) are separated
by $1'.6$ (730 kpc at the cluster redshift) on the sky. The main
physical characteristic of these two subclusters called RX J0152 N and
RX J0152 S are given in table \ref{properties}. A third extended
substructure is also observed at about 1 Mpc to the East of the two
central ones, associated with a galaxy overdensity.

The X-ray analysis\cite{MJE03,HXX04} provides evidence of a possible
merger in progress between the two main subclusters. VLT/FORS
spectroscopic observations\cite{DRL05} have allowed us to confirm the
complex structure observed in the X-rays and giving further support to
the merger hypothesis. The distribution of spectroscopically confirmed
cluster members presents galaxy overdensities which coincide with the
location of the three main X-ray extended regions. The velocity
dispersion of spectroscopic members belonging to RX J0152 N and RX
J0152 S are given in table \ref{properties}. These values, when
compared to the corresponding temperatures of the same substructures,
are in good agreement with the observed $\sigma$-T relation,
indicating that RX J0152 N and S are likely two clusters in a
dynamical state close to virialized. Their masses are estimated to be
$ (2.5 \pm 0.9) \times 10^{14} M_{\odot}$ and $ (1.1 \pm 0.4) \times
10^{14} M_{\odot}$ for RX J0152 N and S respectively.\cite{DRL05}

The main substructures detected in the ICM and galaxy distribution are
also recovered by weak lensing analyses of the cluster,\cite{JWB05} as
shown in Fig. \ref{c0152}-Right, which confirm that most of the baryon
(ICM and galaxies) content is in the potential well created by the DM
distribution. The observed offsets between the centroids of the gas,
galaxy and DM distributions of the three main substructures are
consistent with the picture in which RX J0152 is still in the process
of assembling its mass through the accretion/merging of
substructures.\cite{JWB05} Assuming that the two main sub-clusters are
coming together for the first time and considering their projected
separation and relative velocity, the required time to first crossing
is estimated to be $\Delta t_{c} \simeq 0.4$ Gyr. This time implies
that the virialization of the cluster as a whole will begin at a
redshift $z_c \simeq 0.75$.

Because of the non-uniform distribution of the baryons and DM, the
overall structure of the cluster is likely not to be in virial
equilibrium. Indeed, the overall velocity dispersion of the cluster
turns out to be too high to be consistent with the observed $\sigma$-T
relation.\cite{DRL05} Therefore, an accurate estimate of the virial
mass of the whole cluster, based on the velocity distribution of
spectroscopic cluster members, is not possible.

\subsection{RX J0152: Galaxy populations}

The spectroscopic survey carried out with the VLT allowed us to obtain
spectra for 102 cluster members.\cite{DRL05} We use these spectra to
separate cluster members in two groups: galaxies with emission lines
(star-forming galaxies) and galaxies without emission line features
(passive galaxies). Passive galaxies and star-forming galaxies are
indicated by small circles and triangles respectively in
Fig. \ref{c0152}-Left. One striking feature is that all the
star-forming galaxies are distributed in the outskirts of the cluster,
avoiding the high density regions traced by the X-ray
emission\cite{DRL05,HDR05} (see Fig. \ref{c0152}-Left). The derived
star-formation rate (SFR) of these galaxies\cite{HDR05} turns out to
be smaller than that of galaxies in the field environment. A simple
interpretation for this is that galaxies at large radii from the
cluster center are falling into the cluster potential, suppressing
their star-forming activity as they interact with the ICM (see Homeier
et al. 2005).\cite{HDR05}

An anlysis of the ground based\cite{DRL05} and ACS (Blakeslee et al.,
in prep.) photometry of this cluster shows a clear CM relation in the
corresponding CM diagram. The locus of the CM relation is mainly
defined by passive galaxies, although a few emission line galaxies are
also observed in this region of the diagram. By comparing these data
to models\cite{KA97,BCh93} we find that passive cluster galaxies have
colors consistent with passive evolution of early-type galaxies formed
at $z \simgreat 2$. 

The few emission line galaxies observed within 1$\sigma$ from the best
fit of the locus of the CM relation have colors as red as passive
early-type (elliptical) galaxies. This suggests the existence of an
old stellar population inhabiting these galaxies, although intrinsic
dust extinction may also be the cause of the observed reddening. The
ACS images reveal that most of these galaxies indeed have a red bulge
structure in the center surrounded by a bluer disk, where the
star-forming activity should be taking place.  A more detailed
analysis of the spectroscopic data shows that some of these red
star-forming galaxies are also hosts of a young ($\sim$ 1 Gyr old)
stellar population (see Demarco et al. [2005]\cite{DRL05} for a more
detailed discussion). The fact that these galaxies harbor stellar
population of different ages raises the interesting possibility that
these objects may be undergoing a transition stage between late-type
field galaxies falling into the cluster and early-type passive
galaxies located in the cluster core. Galaxy-ICM interactions may play
a considerable role in this transition phase.

\section{The galaxy cluster RDCS J1252 (z=1.237)}

RDCS J1252 is one of the most X-ray luminous cluster discovered to
date at $z > 1$.\cite{R04b} Its main physical properties are
summarized in table \ref{properties}. Extensive spectroscopy with the
VLT has confirmed its redshift to be $z=1.237$. This cluster is among
the three clusters at $z > 1.2$ observed with the ACS\cite{F04} and is
providing fundamental information to understand cluster formation and
galaxy evolution. The main results to date are reviewed in the
following sections.

\begin{figure}[th]
\centerline{\psfig{file=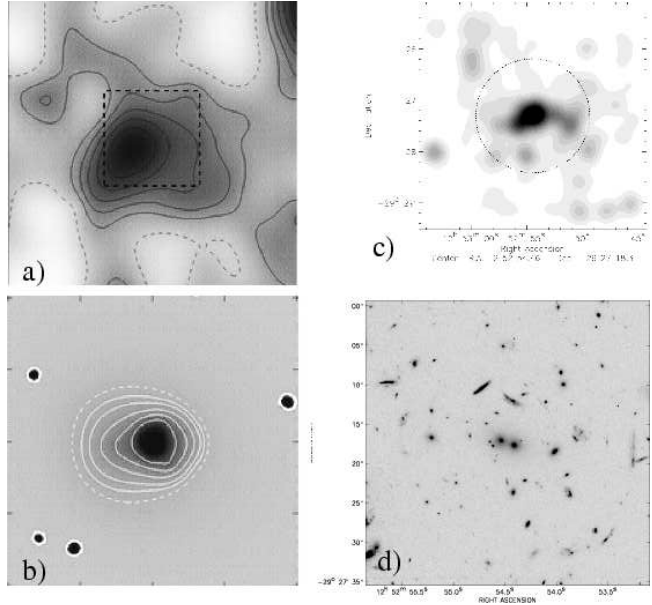,width=8cm,angle=-90}}
\vspace*{8pt}
\caption{RDCS J1252. {\bf a)} ACS weak lensing mass map (taken from
Lombardi et al. 2005) centered on the optical center of the cluster
and covering 5.7 $\times$ 5.7 arcmin$^2$. The dashed square has a side
length of 2 arcmin (1 Mpc at the cluster redshift). {\bf b)} Chandra
X-ray iso-contours in the central 2 arcmin of the cluster (dashed
square; taken from Lombardi et al. 2005). {\bf c)} Map of the smoothed
K-band light of photometric member galaxies. The circle marks the
central 130 arcseconds diameter region of the cluster; taken from Toft
et al. (2004). {\bf d)} The 0.6 $\times$ 0.6 arcmin$^2$ ACS optical
image of the cluster core, dominated by two elliptical galaxies. In
all panels North is up and East is left. The distributions of DM, gas
and cluster members (K-band light) are elongated in the East-West
direction.}
\label{c1252}
\end{figure}

\subsection{RDCS J1252: Cluster structure and dynamics}

The X-ray extended emission of RDCS J1252 is in clear contrast, in
terms of spatial distribution, to that of RX J0152. The X-ray
iso-contours shown in Fig. \ref{c1252}-b are those from Chandra. The
more uniform distribution of the cluster X-ray emission, compared to
that of RX J0152, suggests that the ICM is in a more advanced state of
relaxation, implying that it had to be assembled at higher
redshift. However, when looking at it in more detail, a coma-like
asymmetry in the East-West direction on the sky of the X-ray isophotes
(Fig. \ref{c1252}-b) suggests the possibility that the ICM has been
affected by a recent merging episode. Interestingly, the spatial
distribution of the cluster K-band (2.2 $\mu$m) luminosity weighted
light distribution (Fig. \ref{c1252}-c) from photometric members
(including spectroscopic ones) shows a similar elongated
structure.\cite{DRL04,TMR04} The elongation of the cluster is
confirmed by the weak lensing analysis of the cluster that finds a
similar East-West elongation in the DM distribution\cite{LRB05}
(Fig. \ref{c1252}-a). The cluster core, as seen by the ACS, is shown
in Fig. \ref{c1252}-d.

We derived the cluster luminosity, ICM temperature and metallicity
from the X-ray data\cite{R04b}, while the spectroscopic data allowed
us to securely identify cluster galaxies and thus determine the
cluster velocity dispersion (see table \ref{properties}). The
comparison of these measurements with those of clusters at lower
redshift allow us to study the evolution of scaling relations from
$z=1.24$ down to the local universe. The ICM temperature and velocity
dispersion of RDCS J1252 are in good agreement with the observed
$\sigma-T$ relation while the X-ray luminosity, entropy ($S$) and
temperature values suggest only a mild positive evolution of the
$L_x-T$ and $S-T$ relations.\cite{ETB04} RDCS J1252 appears thus well
thermalized with thermodynamical properties similar to those of
clusters at low redshift. The value of the metallicity obtained from
the XMM spectroscopic data\cite{R04b} turns out to be consistent with
the mean ICM metallicity value for lower redshift
clusters.\cite{TRE03} This result thus provides further support for
the lack of evolution of the amount of metals in galaxy clusters up to
$z \simeq 1.3$ and is consistent with the major episode of metal
enrichment occurring at $\sim 3$. The total mass of the cluster is
estimated to be $(1.9 \pm 0.3) \times 10^{14} M_{\odot}$ within a
radius of 536 kpc,\cite{R04b} consistent with the $\sigma-M$ relation
predicted by Bryan \& Norman (1998).\cite{BN98} In general, the
structure and physical properties of RDCS J1252's ICM show that this
cluster is a massive system in an advanced thermodynamical state, with
scale properties similar to those of clusters at low redshift, that we
are observing at a lookback time of 8.5 Gyrs (when the universe was
about 36\% of its present age).

\subsection{RDCS J1252: Galaxy populations}

So far a total of 38 cluster members have been spectroscopically
confirmed with VLT/FORS (Demarco et al., in prep.). Like RX J0152,
star-forming galaxies are preferentially located in the outer
(infalling) regions of the cluster, with passive early-type galaxies
to dominate the cluster center.

The high angular resolution of the ACS allows us to examine in detail
the morphology of cluster galaxies even at such a large lookback time
(Rosati et al., in preparation). One interesting feature is that,
while all the passive member galaxies are well characterized by an
early-type appearance (bulge-like structure), only one of the
star-forming cluster members shows a well defined morphology with
grand design spiral arms or a disk structure surrounding a central
bulge. Most of these galaxies look quite irregular, suggesting that
they may be still forming. This is in contrast to RX J0152, where many
star-forming galaxies can be clearly classified as spiral
galaxies. More generally speaking, and excluding passive galaxies, the
comparison of galaxy morphologies between RX~J0152 and RDCS J1252
suggests that galaxy formation is still on-going at $z \sim 1.2$ and
that Hubble-type spirals may not be established until $z \simless
1$. However, more observations of high-z galaxies are needed to
strengthen this conclusion.

The combination of ground based spectroscopy with near-IR imaging from
VLT has allowed us to compute the K-band galaxy LF of the
cluster\cite{TMR04} which, at the redshift of $z=1.24$, allows us to
probe the rest-frame $z$-band light from the cluster. When comparing
the derived LF with the $z$-band LF of galaxy clusters in the local
universe a non evolution of the shape of the bright end of the LF is
observed, indicating that massive elliptical galaxies which dominate
the bright end of the galaxy LF were already in place at
$z=1.24$. Another important observation is the brightening of $M^*$
(the characteristic magnitude associated with $L^*$) compared to lower
redshift clusters. This finding is in agreement with the result from
previous work\cite{dPSE99} indicating that galaxies in clusters become
brighter at higher redshifts.

The same ground based near-IR data\cite{LRD04} and ACS optical
data\cite{B03} have been used to study the distribution of the cluster
galaxy populations in Color-Magnitude space. These studies, both
independent, arrive at similar conclusions. A clear and tight CM
relation for early-type (passive) galaxies is observed in RDCS J1252
at $z=1.24$ with a slope and scatter similar to lower redshift
values.\cite{SED98} These results thus confirm that the CM relation of
cluster early-type is a metallicity-mass relation instead of an
age-mass relation, i.e., more massive (luminous) early-type galaxies
are able to better retain metals and thus to produce redder
colors.\cite{KA97} Although the scatter in colors of the early type
galaxies permits considerable variation in the ages, their mean
luminosity weighted age is quite old, corresponding to $z_f \sim 3$.

Both CM studies and the K-band galaxy LF of RDCS J1252 also indicate
that massive early-type galaxies evolved passively in luminosity since
their epoch of formation. The above results for early-type galaxies in
RDCS J1252 are in agreement with studies of early-type galaxies in the
field\cite{P03} and pose challenging problems to hierarchical models
of galaxy formation. However, despite the fact that early-type
galaxies in RDCS J1252 are home for stars with a mean age of
approximately 2.7 Gyrs,\cite{LRD04} the spectroscopic data available
suggest an interesting new result. When combining the spectra of the
10 brightest spectroscopic members, a prominent H$_{\delta}$ feature
emerges.\cite{R04a} This balmer absorption line indicates the presence
of young A- and F-type stars, born about 1 Gyr previous to the epoch
of observation. Although a quantitative analysis is needed to confirm
this result, this observation would indicate that most luminous
early-type galaxies in RDCS J1252 host young (post-starburst) stellar
populations produced in recent episodes of star formation at $z
\simless 2$. We would be thus approaching the formation epoch of these
galaxies as we observe traces of their latest star-forming episode.

\section{Conclusions}

The unprecedented panchromatic study carried out so far on two of the
most distant clusters of galaxies known to date is providing us with
crucial information to better understand cluster evolution and the
formation and evolution of luminous early-type galaxies, the most
massive galaxies known. The main conclusions can be summarized as
follows. Scaling relations such as the $L_x$-T relation and ICM
metallicity do not significantly evolve up to $z=1.24$ pushing energy
injection processes and metal production in galaxy clusters to $z
\simgreat 3$. The analysis of the CM relation in RDCS J1252 at
$z=1.24$ provides valuable information to understand the formation and
evolution of massive elliptical galaxies. These galaxies would have
formed the bulk of their stars at $z \sim 3$ (when the universe was
$\sim$15\% of its present age) evolving passively in luminosity since
then. The large lookback time of formation of massive (luminous) early
type galaxies and their passive evolution are consistent with the
results from galaxy LF studies in cluster and field
environments. Furthermore, these studies support the simple model of
monolithic collapse for the formation of elliptical galaxies. In
particular, the LF of galaxies in the field\cite{P03} suggests that
most of the merging which form elliptical galaxies in the hierarchical
models should occur at redshift $z > 2-3$. This study poses
difficulties for hierarchical models which underpredict the density of
luminous galaxies at $z \geq 1$ and overpredict significantly the
density of low luminosity galaxies at $z \leq 1$.

In addition to all the above, the joint analysis of the available data
on all the clusters observed with the ACS is providing more robust and
statistically significant information about the structure and
evolution of clusters and cluster galaxies. In particular, the
morphology-density relation\cite{D80,GYF03} is well established at $z
\sim 1$, an epoch when the universe was $\sim$40 \% its present
age. Late-type star-forming galaxies are observed to prefer the outer
low density regions of clusters, while early-type passive galaxies
populate the high density regions of the cluster cores. Interesting
evolutionary trends are observed in the morphology-density relation as
a function of density and redshift\cite{PFC05}. The available data
suggest that this evolution is primarily driven by evolution in the
fraction of S0 and Spiral+Irregular galaxies: a deficit of S0s and an
excess of Spiral+Irregular galaxies are observed at $z \sim 1$
relative to the local galaxy population.\cite{PFC05} Moreover, mild
correlations of the cluster early-type fraction and cluster
star-formation rate with cluster X-ray luminosity have been
detected.\cite{PFC05,HDR05}

In the future, the integration of the Spitzer/IRAC data available on
the clusters in the ACS cluster program will provide new information
about the bulk of the stellar mass in galaxies (whose light peaks at
about 2 $\mu$m rest frame) at redshifts $z \simgreat 1$. Also, the
strong lensing analysis of some of the clusters observed with the ACS
will allow us to accurately model the mass distribution in the central
regions of distant clusters, thus providing important tests of the
Cold Dark Matter (CDM) paradigm. In addition to the recent discovery
of an X-ray luminous cluster at $z=1.4$,\cite{MRL05} more clusters at
similar and higher redshift need to be discovered in order to bridge
the evolutionary gap between our most distant clusters and their
likely progenitors, the so called proto-clusters which have been
discovered at redshifts $z \geq 2$.\cite{V02,M04}

\section*{Acknowledgments}

ACS was developed under NASA contract NAS5-32865. We thank our fellow
ACS team members for their important contributions to this research
and we also thank the support from ESO staff in Chile and Germany. We
are grateful to Ken Anderson, John McCann, Sharon Bushing, Alex
Framarini, Sharon Barkhouser, and Terry Allen for their invaluable
contributions to the ACS project at JHU. This investigation has been
partially supported by NASA grant NAG5-7697.

\end{document}